\documentclass[12pt]{article}
\usepackage{amsmath}
\usepackage{latexsym}

\begin{document}
\renewcommand{\thefootnote}{\fnsymbol{footnote}}
\null\vskip-24pt \hfill KL-TH/03-04  \vskip0.3truecm
\begin{center}
\vskip 3truecm {\Large\bf
On the supermultiplet of anomalous currents  in d=6}\\
\vskip 1.5truecm
{\large\bf R. Manvelyan\footnote{On leave from Yerevan Physics
Institute, email: manvel@moon.yerphi.am}, W. R\"uhl}\\
\vskip 1 truecm
{\it Department of Physics, Theoretical Physics\\
University of Kaiserslautern, P.O.Box 3049 \\
67653 Kaiserslautern, Germany}\\
\medskip
{\small\tt manvel,ruehl@physik.uni-kl.de}
\end{center}

\vskip 3truecm \centerline{\bf Abstract} The multiplet of
superconformal anomalous currents  in the case $(1,0)$, \\$d=6$
is derived. The  supersymmetric multiplet of  anomalies contains
the trace of the energy momentum tensor, the gamma trace of the
supercurrent and some topological vector current with divergence
equal to the $R$-current anomaly. The extension of this
consideration to the $(2,0)$ case and some application and
motivation coming from $AdS_{7}/CFT_{6}$ correspondence is
discussed.
\newpage
\vspace*{-2cm}
\section{Introduction}
\renewcommand{\thefootnote}{\arabic{footnote}} \setcounter{footnote}0
AdS/CFT correspondence \cite{Malda} provides a new tool for
understanding the strong coupling dynamics of gauge theories. The
most developed example is IIB string/gravity on $AdS_{5}\times
S^5$ describing the large N dynamics of ${\cal N} =4$
Super-Yang-Mills theory . The investigation of the anomalies
\cite{anom} and the correlation functions in this approach
\cite{d4, HS} is very important because their universal behavior
with respect to the renormalization from the weak to the strong
coupling regime \cite{petsken} offers a unique possibility to test
this correspondence using the free field approach in the gauge
theory calculations. Another interesting example of this
correspondence is  $AdS_{7}/CFT_{6}$ case describing duality
between the M theory/11dSUGRA and the maximal supersymmetric (2,0)
tensor multiplet in d=6 \cite{d6}. This theory deals with the low
energy regime of N coincident M5-branes and has some mysterious
properties like the absence of a free coupling parameter, the
difficulties in the lagrangian formulation of the self-dual
antisymmetric tensor field and its nonabelian generalization. The
test of $AdS_{7}/CFT_{6}$ includes the investigation of two- and
three-point correlation functions of the energy-momentum tensor
and vector R-currents \cite{BFT1, ManPet1}, renormalization
properties of trace anomaly coefficients in external gravitational
and  vector fields \cite{BFT2, ManPet2} and the behavior of the
R-current anomaly from the weak to the strong coupling limit
\cite{minasian}. These investigations lead to some discrepancy in
contrast to the $AdS_{5}/CFT_{4}$ case concerning the behavior of
the coefficient of the Euler density term of the trace anomaly
\cite{BFT2, arkady} and the general structure of the R-anomaly in
different energy limits \cite{minasian}. These phenomena still
need further explanations. In this note we investigate the SUSY
structure of the superconformal current multiplet in $d=6$ in the
(1,0) case and discuss the (2,0) current and anomaly multiplets
also. Our goal is to construct the multiplet of anomalies
including the trace and the R-current anomalies as bosonic
components in $d=6$ using \emph{rigid} SUSY algebra. This can help
to understand the strange behavior of the anomalous coefficients
during renormalization from the weak (free fields) to strong
(AdS/CFT) coupling. The result described below is the following.
\emph{The anomaly multiplet in $d=6$ includes in one superfield
the trace anomaly of the energy-momentum tensor, the gamma-trace
of the supersymmetry current and the topological Chern-Simons
current whose divergence is equal to the R-current anomaly.}
\section{Superconformal currents in the (1,0) case and the anomaly multiplet}
The conserved conformal supercurrent in the case $d=6$, $(1,0)$
case forms the scalar superfield $J$ satisfying the third order
conservation condition \footnote{Our conventions and notations can
be found in the Appendix} \cite{HST}
\begin{equation}\label{CE}
D^{3\alpha ijk}J=\epsilon^{\alpha\beta\gamma\delta}
D^{(i}_{\beta}D^{j}_{\gamma}D^{k)}_{\delta}J=0
\end{equation}
From this equation and the condition of closure of the SUSY
algebra one can derive (see Ref.\cite{HST}) the independent
components, their transformation and conservation  rules. Here we
will only list the independent components in ascending order of
Weyl weights (dimensions) and present  the conservation
conditions.
\begin{eqnarray}\label{cm}
\begin{array}{|c|c|c|} \hline
 J & \textnormal{off-shell scalar}& \textnormal{auxiliary field} \\
\hline
 \psi^{i}_{\alpha} & \textnormal{off-shell fermion}
& \textnormal{auxiliary field} \\
\hline
 C_{\alpha\beta} & C_{\alpha\beta}=C_{(\alpha\beta)}  &
 \textnormal{off-shell auxiliary self-dual} \\
   & & \textnormal{third rank tensor}  \\
\hline
 V^{ij}_{\alpha\beta} & V^{ij}_{\alpha\beta}= V^{(ij)}_{[\alpha\beta]},
 & \textnormal{conserved SU(2)-triplet}  \\
   & \partial^{\alpha\beta}V^{ij}_{\alpha\beta}=0 &
\textnormal{R-symmetry current}  \\
\hline
 S^{i}_{\gamma,\beta\alpha} & S^{i}_{\gamma,\beta\alpha}=S^{i}_{\gamma,[\beta\alpha]}
 , S^{i}_{[\gamma,\beta\alpha]}=\Sigma^{a \alpha\beta}S^{i}_{a,\beta}=0  & \textnormal{conformal, conserved}  \\
  & \partial^{\beta\alpha}S^{i}_{\gamma,\beta\alpha}=0 & \textnormal{supersymmetry current}\\
\hline
 T_{\delta\gamma,\beta\alpha} &T_{\delta\gamma,\beta\alpha}=T_{[\delta\gamma],[\beta\alpha]}
 =T_{\beta\alpha,\delta\gamma}, &
 \textnormal{conformal, conserved} \\
  &  \partial^{\beta\alpha}T_{\beta\alpha,\delta\gamma}=0 ,
  T_{\delta[\gamma,\beta\alpha]}=T^{\quad \beta\alpha }_{\beta\alpha}=0 & \textnormal{energy-momentum tensor}\\
\hline
\end{array}
\end{eqnarray}

We want to deform (\ref{CE}) with some off-shell supermultiplet of
anomalies
\begin{equation}\label{an}
 D^{3\alpha ijk}J=A^{\alpha ijk} ,
\end{equation}
but with structure maintaining  conservation of the supersymmetry
current and energy-momentum tensor
\begin{equation}\label{st}
\partial^{\beta\alpha}S^{i}_{\gamma,\beta\alpha}=0 \quad ,
\quad\partial^{\beta\alpha}T_{\beta\alpha,\delta\gamma}=0
\end{equation}
and violating the tracelessness and gamma tracelessness
conditions of the latter and the conservation of the vector
$SU(2)$ R-current
\begin{eqnarray}\label{Trace}
    \epsilon^{\alpha\beta\gamma\delta}S^{i}_{\beta,\gamma\delta}\neq 0\quad,
    \quad T^{\quad\alpha\beta}_{\alpha\beta}\neq 0\quad ,\quad
     \partial^{\alpha\beta}V^{ij}_{\alpha\beta}\neq 0
\end{eqnarray}
 More precisely, we want to
find a superfield  with the last two levels (with Weyl weight 11/2
and 6) of independent components
\begin{equation}\label{an1}
    \dots ; \quad\xi^{i\alpha}\quad ; \quad \Theta \quad , \quad
    v^{ij}=v^{(ij)}
\end{equation}
which we can use for parametrization of the general anomaly
superfield (\ref{an}) and which leads to the violation of the
conservation conditions (\ref{Trace})(the analogous  consideration
for the gauge and gravitational anomalies in (1,0) d=6 case can be
found in Ref. \cite{HU}). First of all we have to write a possible
transformation of the set of fields (\ref{an1})
\begin{eqnarray}
  D^{j}_{\beta}\xi^{i\alpha} &=&
  \delta^{\alpha}_{\beta}\left(\varepsilon^{ji}\Theta + v^{ji}\right)
  + M^{\alpha}_{\beta}\varepsilon^{ji} + M^{\alpha (ji)}_{\beta} +
   \dots \quad , \label{dots}\\
   D^{i}_{\alpha}\Theta &=& -\frac{a}{2}
   i\partial_{\alpha\beta}\xi^{i\beta} ,\label{dt}\\
 D^{k}_{\alpha}v^{ji}&=&-b\varepsilon^{k(j}i\partial_{\alpha\beta}\xi^{i)\beta} ,\label{dv}\\
M^{\alpha}_{\alpha}&=&M^{\alpha (ji)}_{\alpha}=0 .\label{tm}
\end{eqnarray}
here dots in (\ref{dots}) replace derivative terms of  unknown
auxiliary fields with lower Weyl weights. The tracelessness
conditions (\ref{tm}) for the general antisymmetric and symmetric
part of transformation of $\xi^{i\alpha}$ are natural because we
already extracted the $\delta^{\alpha}_{\beta}$ trace of
$D^{j}_{\beta}\xi^{i\alpha}$ in (\ref{dots}) as a bracket with our
anomalies. Note that the presence in r.h.s. of equations
(\ref{dt}) and (\ref{dv}) of   only the derivatives of the lower
spinor components expresses the lack of  independent components of
the anomaly multiplet with  higher Weyl weights. Then we can try
to fix the unknown coefficients $a, b$ and possible configurations
for the second rank antisymmetric tensor fields
$M^{\alpha}_{\beta}$ and $M^{\alpha (ij)}_{\beta}$. Immediately we
obtain the following restrictions
\begin{eqnarray}
 &&\left\{D^{i}_{\alpha}, D^{j}_{\beta}\right\}\Theta =
i\varepsilon^{ij}\partial_{\alpha\beta}\Theta \quad \Rightarrow
\partial_{\lambda(\alpha}M^{\lambda (ij)}_{\beta)} = 0 \quad ,
  \quad   \partial_{\lambda[\alpha}M^{\lambda}_{\beta]} = 0\label{ct}\\
 &&\left\{D^{k}_{\alpha}, D^{l}_{\beta}\right\}v^{ji}=
i\varepsilon^{kl}\partial_{\alpha\beta}v^{ji}\Rightarrow\partial_{\lambda(\alpha}
M^{\lambda}_{\beta)} = 0 \quad ,
  \quad   \partial_{\lambda[\alpha}M^{\lambda (ij)}_{\beta]} = 0\label{cv}
\end{eqnarray}
So we obtain for both tensor fields not only a Bianchi  identity
(symmetric second rank equations) leading to the corresponding
vector field potentials, but get another two restrictions
(antisymmetric second rank equations) leading to the on-shell
conditions for these potentials. So we prove the statement:
\emph{There is no off-shell supermultiplet of fields containing
the set of anomalies (\ref{an1}) as highest Weyl weight
independent components. }

Nevertheless we will try to construct the analogy of  the well
known  relation in the case $N=1$, $d=4$  for the anomalous
currents
\begin{equation}\label{4d}
    \bar{D}^{\dot{\alpha}}J_{\alpha\dot{\alpha}}=D_{\alpha}A, \quad
\bar{D}_{\dot{\alpha}}A=0
\end{equation}
looking for the possible multiplets with off-shell and absolutely
unrestricted components (even without Bianchi identities or
conservation condition). This type of multiplet could play the
role of the chiral multiplet in $d=4$ (\ref{4d}) expressing
anomalies in N=1, and also in N=2 cases. But in $d=6$ there is no
chiral multiplet, instead we  have here \emph{only one known
irreducible supermultiplet with unrestricted independent
components}. It is the $SU(5)$ \textbf{5} superfield
$L^{ijkl}=L^{(ijkl)}$ constrained with the following superfield
condition
\begin{equation}\label{5plet}
   D^{(i}_{\alpha}L^{jklm)}=0
\end{equation}
but containing only the off-shell and  unconstrained  components
\begin{eqnarray}
 L^{ijkl} \quad &,& \quad  \lambda^{jkl}_{\alpha}=D_{\alpha i}L^{ijkl}\quad
 ,\nonumber\\
 G^{kl}_{\beta\alpha}=D_{\beta j}\lambda^{jkl}_{\alpha}\quad ,\quad
 \xi^{i\delta}&=&\epsilon^{\gamma\beta\alpha\delta}D_{\gamma
 j}G^{ji}_{\beta\alpha} \quad ,\quad \Theta=D_{\alpha i}\xi^{i\alpha}\label{5p}
\end{eqnarray}
This multiplet was considered in Ref. \cite{HST}, \cite{K} and
used in Ref. \cite{HU} for the construction of gauge anomalies in
the $(1,0)$ case (the so-called relaxed hypermultiplet
\cite{HStT}). The same superfield we can use for the quantum
deformation of the conservation relation (\ref{CE}) in the
following way
\begin{equation}\label{AE}
D^{3\alpha
ijk}J=-2\epsilon^{\alpha\gamma\beta\delta}i\partial_{\gamma\beta}\lambda^{ijk}_{\delta}
=4i\partial^{\beta\alpha}D_{\beta l}L^{lijk}
\end{equation}
So we express the general anomaly in (\ref{an}) by mixing the
conformal current multiplet (\ref{cm}) with the $SU(2)$
$\textbf{5}$ superfield (\ref{5p}).
\begin{equation}\label{an3}
    A^{\alpha ijk}=4i\partial^{\beta\alpha}D_{\beta l}L^{lijk}
\end{equation}
Note that this equality itself concerns the usual derivative and
does not lead directly to the relation between the traces of the
current multiplet and components of $L^{ijkl}$ (such as in d=4
case (\ref{4d})). However  exploring the relation (\ref{AE}) and
the condition of closure of the supersymmetry  we obtain the
following relations for the transformations of the components and
the set of conservation conditions:

\begin{eqnarray}
 &&\qquad J \qquad ,  \qquad  L^{ijkl} ;\\
 && D^{i}_{\alpha}J =\psi^{i}_{\alpha} \quad , \quad D^{n}_{\alpha}L^{ijkl}=
  \frac{4}{5}\varepsilon^{n(i}\lambda^{jkl)}_{\alpha} ;\\
 && D^{j}_{\beta}\psi^{i}_{\alpha}= V^{ji}_{\beta\alpha} + \varepsilon^{ji}C_{\beta\alpha}
 + \frac{i}{2}\varepsilon^{ji}\partial_{\beta\alpha}J,\quad
  D^{i}_{\beta}\lambda^{jkl}_{\alpha}=
  \frac{3}{4}\varepsilon^{i(j}G^{kl)}_{\beta\alpha}-
  \frac{5}{4} \partial_{\beta\alpha} L^{ijkl} ;\\
 && D^{k}_{\gamma}V^{ji}_{\beta\alpha}=\varepsilon^{k(j}S^{i)}_{\gamma,\beta\alpha}
 -\frac{1}{9}\epsilon_{\gamma\beta\alpha\delta}\varepsilon^{k(j}\xi^{i)\delta}
 -2i\partial_{[\gamma\beta}\lambda^{kji}_{\alpha]} +
 \frac{4}{5}i\varepsilon^{k(j}\partial_{\gamma[\beta}\psi^{i)}_{\alpha]}
 +\frac{1}{5}i\varepsilon^{k(j}\partial_{\beta\alpha}\psi^{i)}_{\gamma}
 ,\qquad\nonumber\\
 && D^{i}_{\gamma}C_{\beta\alpha}= S^{i}_{(\alpha,\beta)\gamma} -
 \frac{4}{5}i\partial_{\gamma(\beta}\psi^{i}_{\alpha)} \quad ,\quad
\begin{array}{|c|}
\hline
 \partial^{\beta\alpha}S^{i}_{\gamma,\beta\alpha}=0 \\
\hline
\end{array}\quad
 ,\quad
 \begin{array}{|c|}
\hline
  S^{i}_{[\gamma,\beta\alpha]}=\frac{1}{9}\epsilon_{\gamma\beta\alpha\delta}\xi^{\delta i}\\
\hline
\end{array}
\quad  ,\\
&& D^{k}_{\gamma}G^{ji}_{\beta\alpha}=
\frac{1}{9}\epsilon_{\gamma\beta\alpha\delta}
\varepsilon^{k(j}\xi^{i)\delta}-
\frac{1}{3}i\partial_{\beta\alpha}\lambda^{kji}_{\gamma}
-\frac{8}{3}i\partial_{\gamma[\beta}\lambda^{kji}_{\alpha]}\quad ;\nonumber\\
&&D^{j}_{\delta}S^{i}_{\gamma,\beta\alpha}=
\frac{1}{2}\varepsilon^{ji}T_{\delta\gamma,\beta\alpha}
+\frac{2}{3}i\partial_{\delta\gamma}V^{ji}_{\alpha\beta}
+\frac{2}{15}i\partial_{\beta\alpha}V^{ji}_{\delta\gamma}
+\frac{2}{15}i\partial_{\gamma[\beta}V^{ji}_{\alpha]\delta}
+\frac{2}{3}i\partial_{\delta[\beta}V^{ji}_{\alpha]\gamma}\nonumber\\
&&\qquad \qquad
+\frac{2}{3}i\partial_{\beta\alpha}G^{ji}_{\gamma\delta}
+\frac{2}{3}i\partial_{\gamma\delta}G^{ji}_{\beta\alpha}
+\frac{1}{3}i\partial_{\gamma[\beta}G^{ji}_{\alpha]\delta}
-\frac{7}{3}i\partial_{\delta[\beta}G^{ji}_{\alpha]\gamma}\\
&&\quad +i\varepsilon^{ji}\partial_{\delta[\alpha}C_{\beta]\gamma}
-\frac{1}{5}i\varepsilon^{ji}\partial_{\gamma[\alpha}C_{\beta]\delta}
-\frac{1}{5}i\varepsilon^{ji}\partial_{\beta\alpha}C_{\delta\gamma}\quad
, \quad
\begin{array}{|c|}
\hline
  \partial^{\beta\alpha}V^{ji}_{\beta\alpha}+\partial^{\beta\alpha}G^{ji}_{\beta\alpha}=0\\
\hline
\end{array}
 \quad ,\nonumber\\
&&
D^{j}_{\beta}\xi^{i\alpha}=\frac{1}{8}\varepsilon^{ji}\delta^{\alpha}_{\beta}\Theta
+\frac{9}{2}i\partial^{\gamma\delta}G^{ji}_{\gamma\delta}\delta^{\alpha}_{\beta}
-6i\partial^{\gamma\alpha}G^{ji}_{\gamma\beta} \quad , \quad
\begin{array}{|c|}
\hline
 \Theta + 3T^{\quad\beta\alpha}_{\beta\alpha}=0 \\
\hline
\end{array}
 \quad ;\nonumber\\
&& D^{i}_{\varepsilon}T_{\delta\gamma,\beta\alpha}=
-\frac{2}{3}i\partial_{[\delta[\alpha}S^{i}_{\varepsilon,\beta]\gamma]}
-2i\partial_{\varepsilon[\delta}S^{i}_{\gamma],\beta\alpha}
-2i\partial_{\varepsilon[\beta}S^{i}_{\alpha],\delta\gamma}
-\frac{1}{3}i\partial_{\delta\gamma}S^{i}_{\varepsilon,\beta\alpha}
-\frac{1}{3}i\partial_{\beta\alpha}S^{i}_{\varepsilon,\delta\gamma},\quad \,\,\\
&& D^{i}_{\alpha}\Theta=-4i\partial_{\alpha\beta}\xi^{i\beta}\quad
, \quad
\begin{array}{|c|}
\hline
 \partial^{\beta\alpha}T_{\beta\alpha,\gamma\delta}=0  \\
\hline
\end{array}
\quad .\nonumber
\end{eqnarray}
where we mark the important  conservation conditions and the
relations between the previously independent components by using
frames. Thus we obtain the following result: \emph{The multiplet
of anomalies in the six dimensional case includes the trace of the
energy-momentum tensor ($\Theta$), the gamma-trace of the
supercurrent ($\xi^{i\alpha}$) and some $SU(2)$ triplet of the
vector currents $G^{ij}_{\alpha\beta}$
 restricted to express the R-symmetry anomaly by it's
divergence}. But the R-symmetry anomaly is kind of the chiral
anomalies and hence it is some linear combination of parity-odd
topological invariants constructed from the external fields.
Therefore we can always express this type of anomaly in the form
of the divergence of some topological  vector currents dual to the
Chern-Simons forms in one dimension less.
\begin{equation}\label{CS}
    {}^*(F\wedge F \wedge F)\sim
    \partial_{a}tr(\epsilon^{abcdef}A_{b}F_{cd}F_{ef}+.....)
\end{equation}
Changing to the usual notations we can say that the following
objects are the components of the single superfield
\begin{eqnarray}
 \Theta &=& D^{4}_{ijkl}L^{ijkl}=12T^{a}_{a}\quad , \label{a1}\\
 \xi^{i\alpha}&=&D^{3\alpha}_{jkl}L^{ijk}= 3S_{\beta,a}\Sigma^{a\beta\alpha}\quad ,\label{a2}\\
 G_{\alpha\beta}^{ij}&=& D^{2}_{\alpha\beta, kl}L^{ijkl}\quad , \quad \textnormal{with} \quad
 \partial^{\alpha\beta}G^{ij}_{\alpha\beta}=4\partial^{a}V^{ij}_{a}\label{a3}
\end{eqnarray}
Of course these relations are dependent on the initial
normalization of the currents in the superconformal current
multiplet (\ref{cm}).

\section{Discussion of superconformal currents and anomalies in (2,0) case}
First of all we note that the previous statement about the absence
in the case of six dimensional (1,0) of an off-shell
supermultiplet with highest independent weight components
corresponding to the trace and R-symmetry anomalies is valid in
the (2,0) case also. We have just to replace all  $SU(2)$
R-symmetry indices with $USp(4)$ and consider instead of
$M^{\alpha}_{\beta}$ the antisymmetric tensor $M^{\alpha
[ij]}_{\beta}$. We will come again to the on-shell condition from
the closure of the SUSY algebra. Nevertheless in this case the
supermultiplet of conserved conformal currents exists and is
described  again by the scalar superfield but now the
$\textbf{14}$ of the $USp(4)$  $J^{ij,kl}$, with the following set
of properties and constraints
\begin{eqnarray}
  J^{ij,kl}&=&J^{[ij],[kl]}=J^{kl,ij}\quad ,
  \quad J^{ij,kl}\Omega_{ij}=0 \quad , \quad J^{ij,kl}\epsilon_{ijkl}=0 \label{js}\\
 D^{m}_{\alpha}J^{ij,kl}&=&\Omega^{m[i}
 \lambda^{j],kl}_{\alpha} +
 \frac{1}{4}\Omega^{ij}\lambda^{m,kl}_{\alpha}
 +\Omega^{m[k}
 \lambda^{l],ij}_{\alpha} + \frac{1}{4}\Omega^{kl}\lambda^{m,ij}_{\alpha} \label{ccond}\\
\lambda^{m,kl}_{\alpha}&=&\lambda^{m,[kl]}_{\alpha} \quad ,\quad
\lambda^{m,kl}_{\alpha}\Omega_{kl}=0 \quad ,\quad
\lambda^{m,kl}_{\alpha}\Omega_{mk}=0 \label{ls}
\end{eqnarray}The components of this  conserved superconformal current multiplet
of $(2,0)$ theory were listed for the first time in Ref.
\cite{HST}

\begin{eqnarray}\label{cm2}
\begin{array}{|c|c|c|} \hline
 J^{ij,kl} & \textnormal{\textbf{14} of $USp(4)$ off-shell scalars see (\ref{js})}& \textnormal{auxiliary field} \\
\hline
 \lambda^{i,jk}_{\alpha} &\textnormal{ \textbf{16} of $USp(4)$ off-shell fermion see (\ref{ls}) }  & \textnormal{auxiliary field} \\
\hline
 C_{\alpha\beta}^{ij} & C^{ij}_{\alpha\beta}=C^{[ij]}_{(\alpha\beta)}  &
 \textnormal{off-shell auxiliary self-dual} \\
   &C_{\alpha\beta}^{ij}\Omega_{ij}=0 & \textnormal{third rank tensor}  \\
\hline
 V^{ij}_{\alpha\beta} & V^{ij}_{\alpha\beta}= V^{(ij)}_{[\alpha\beta]},
 & \textnormal{conserved  \textbf{10} of USp(4)}  \\
   & \partial^{\alpha\beta}V^{ij}_{\alpha\beta}=0 & \textnormal{R-symmetry current}  \\
\hline
 S^{i}_{\gamma,\beta\alpha} & S^{i}_{\gamma,\beta\alpha}=S^{i}_{\gamma,[\beta\alpha]}
 , S^{i}_{[\gamma,\beta\alpha]}=\Sigma^{a \alpha\beta}S^{i}_{a,\beta}=0  & \textnormal{conformal, conserved}  \\
  & \partial^{\beta\alpha}S^{i}_{\gamma,\beta\alpha}=0 & \textnormal{supersymmetry current}\\
\hline
 T_{\delta\gamma,\beta\alpha} &T_{\delta\gamma,\beta\alpha}=T_{[\delta\gamma],[\beta\alpha]}
 =T_{\beta\alpha,\delta\gamma}, &
 \textnormal{conformal, conserved} \\
  &  \partial^{\beta\alpha}T_{\beta\alpha,\delta\gamma}=0 ,
  T_{\delta[\gamma,\beta\alpha]}=T^{\quad \beta\alpha }_{\beta\alpha}=0 & \textnormal{energy-momentum tensor}\\
\hline
\end{array}
\end{eqnarray}
and the transformation rules and the coupling with conformal
supergravity were considered in Ref. \cite{BSP}. Following the
previous section we could try to deform the superfield
conservation condition (\ref{ccond}) introducing the superfield
$A^{m,[ij],[kl]}_{\alpha}$ expressed through some \emph{off-shell}
multiplet of anomalies containing a topological Chern-Simons
current $G^{(ij)}_{[\alpha\beta]}$ ,
$\partial^{\alpha\beta}G^{ij}_{\alpha\beta}\sim
\partial^{\alpha\beta}V^{(ij)}_{\alpha\beta}$ and the supersymmetry
and the trace anomalies $\xi^{i\alpha}$ and $\Theta$.
Unfortunately the structure of the off-shell multiplets in the
maximal extended (2,0) case is not well understood  and we can not
present here the full solution of this problem, but a partial
answer is the following.
\begin{eqnarray}
  D^{i}_{\alpha}\Theta &=& -\frac{1}{2}i\partial_{\alpha\beta}\xi^{i\beta}
  \quad , \label{trace}\\
  D^{j}_{\beta}\xi^{i\alpha}&=&\delta^{\alpha}_{\beta}\Omega^{ji}\Theta
-\frac{6}{5}i\partial_{\beta\gamma}G^{\alpha\gamma ji}-
\frac{1}{5}\delta^{\alpha}_{\beta}i\partial_{\gamma\delta}G^{\gamma\delta
ji}-
\frac{4}{5}i\partial_{\beta\gamma}C^{\alpha\gamma ji} \label{susyc}\\
 G^{\alpha\gamma ji}&=& G^{[\alpha\gamma] (ji)} \quad , \quad  C^{\alpha\gamma ji}
=C^{(\alpha\gamma) [ji]} \quad ,\quad C^{\alpha\gamma ji}\Omega_{ji}=0 \nonumber \\
 D^{k}_{\gamma}G^{\beta\alpha ij}&=&
\delta^{[\beta}_{\gamma}\xi^{\alpha](i}\Omega^{j)k}+ \dots \label{g}\\
D^{k}_{\gamma}C^{\beta\alpha ij}&=&
\delta^{(\beta}_{\gamma}\xi^{\alpha)[j}\Omega^{i]k}+
\frac{1}{4}\delta^{(\beta}_{\alpha}\xi^{\alpha)k}\Omega^{ji} +
\dots \label{c}
\end{eqnarray}
Note that the coefficients in  (\ref{trace}) and (\ref{susyc}) and
the presence of the new auxiliary tensor field
$C^{\alpha\beta[ij]}$ are necessary  from the condition of the
closure of supersymmetry algebra on $\Theta$ and $\xi^{i\alpha}$.
Unfortunately we can not present the complete  tower of auxiliary
fields \footnote{the dots in (\ref{g}) and (\ref{c}) mean the
possible derivatives of the next fermion field
$\psi^{i[jk],\gamma\beta\alpha}$}, but can predict that if this
supermultiplet of anomalies exists, the relation between different
anomalies has to remain in the same way as in
(\ref{a1})-(\ref{a3}). For comparison let us remember the $d=4$
$N=2$ and $N=1$ cases. In the $N=2$ case the conformal conserved
currents and anomalies described by just a scalar real superfield
and $N=2$ chiral multiplet correspondingly (see e.g. \cite{HW} and
ref. there) with the following anomalous relation
\begin{eqnarray}
 D^{ij}J &=& \bar{D}^{ij}S \quad , \bar{D}^{i}_{\dot{\alpha}}S=0 \\
 D^{ij}&=&D_{\alpha}^{i}D^{j\alpha}\nonumber
\end{eqnarray}
The $N=2$ anomaly multiplet $S$ is in the language of $N=1$
superfields equivalent to the chiral scalar and vector multiplet.
So the  $N=1$ anomaly multiplet $A$ in Eq. (\ref{4d}) can be
identified  with this scalar chiral part of the $N=2$ chiral
superfield. It means that the trace anomaly and $U(1)$ part of
$SU(2)$ R-symmetry anomaly of the $N=2$ theory form this $N=1$
superfield with the same relative coefficients, and extension of
SUSY just adds a deformation to the conservation laws  of
additional components of supercurrent and R-current. Of course as
an artefact of this splitting to $N=1$ superfields we will get the
Konishi anomaly.

Something like this we can expect in six dimensions. It means that
our (1,0) anomaly multiplet $L^{ijkl}$ could be some part of an
unknown (2,0) anomaly multiplet superfield if we will consider
that in the language of the (1,0) superfields. This consideration
is in progress and will be studied elsewhere. The concrete
application of our result to the explanation of the behavior of
the anomaly coefficients of $(2,0)$ multiplet also needs
additional considerations.  Another interesting task connected
with the latter is to express the anomaly superfield through the
linearized external conformal supergravity (so-called Weyl
multiplet) and find the superfield generalization of the Euler
density and the Weyl invariant combination of curvature
\cite{anom} which in $d=6$ possibly contribute to the conformal
anomaly.

 {\bf Acknowledgement}

Work of R.M.  supported by DFG (Deutsche Forschungsgemeinschaft)
and partially supported  by the Volkswagen Foundation of Germany.
The authors would like to thank R. Mkrtchyan, R. Minasian and T.
Leonhardt for discussions.
\section{Appendix \\Notation and Convention}
\setcounter{equation}{0}
\renewcommand{\theequation}{A.\arabic{equation}}

We use the normalized antisymmetrization $[\dots]$ and
symmetrization $(\dots)$ of all type of indices. Latin indices
$a,b,c,...=1,2,..6$ we use for usual $SO(1,5)$ tensors and Greek
indices  $\alpha, \beta,\gamma,...=1,2,3,4$ for spinor $SU(4)$.
The convention for  $d=6$ symplectic Majorana-Weyl spinors is
\cite{KT}:
\begin{equation}\label{SMW}
    \left(\psi^{i}_{\alpha}\right)^{*}\equiv
    \Omega_{ij}B^{\beta}_{\dot{\alpha}}\psi^{j}_{\beta}, \quad
    B^{*}B=-1
\end{equation}
Here in the $(1,0)$ case $R$-symmetry group indices $i,j,k,..$
belong to $SU(2)$, (i,j,..=1,2) and the corresponding invariant
symplectic metric $\Omega_{ij}=\varepsilon_{ij}$. In the $(2,0)$
case the corresponding $R$-symmetry group is $USp(4)$,
(i,j,..=1,2,3,4) and we have two symplectic tensors $\Omega_{ij}$
and $\epsilon_{ijkl}=3\Omega_{i[j}\Omega_{kl]}$. But in both cases
the existence of the unitary matrix $B$ (\ref{SMW}) which
transfers dotted and undotted indices allows us to formulate
everything without dotted $SU(4)^{*}$ indices using only $d=6$
chiral  SUSY algebra
\begin{eqnarray}\label{SUSY}
    \left\{D^{i}_{\alpha},
    D^{j}_{\beta}\right\}&=&i\Omega^{ij}\partial_{\alpha\beta}\quad
    \quad (\Omega^{ij}=\varepsilon^{ij} \quad\textnormal{in the (1,0)
    case})
\end{eqnarray}

$\partial_{\alpha\beta}=\Sigma^{a}_{\alpha\beta}\partial_{a}$ and
$\Sigma^{a}_{\alpha\beta}$ is the set of 6 antisymmetric sigma
matrices coming from the definition of usual $d=6$ gamma matrices
in the following way with the corresponding normalization and
completeness rules \cite{HST}
\begin{eqnarray}
  &&\Sigma^{a}_{\alpha\beta}= -B^{\dot{\beta}}_{\beta}
  \Sigma^{a}_{\alpha\dot{\beta}}=-\Sigma^{a}_{\beta\alpha} \\
 &&\Gamma^{a}=\left(\begin{array}{cc}
  0& \Sigma^{a}_{\alpha\dot{\beta}}\\
  \tilde{\Sigma}^{a\dot{\alpha}\beta}& 0
\end{array} \right)\\
&&\Sigma^{a}_{\alpha\beta}\Sigma^{b\alpha\beta}=-4\eta^{ab}\quad , \quad \eta^{ab}=(+,-----) \\
&&\Sigma^{a}_{\alpha\beta}\Sigma_{a\gamma\delta}=-2\epsilon_{\alpha\beta\gamma\delta}
\end{eqnarray}
Following this definition we can present a table of conversion and
rules for corresponding tensors and spin-tensors.
$$\begin{array}{|c|c|} \hline
 V_{a} & V_{\alpha\beta}=-V_{\beta\alpha}  \\
\hline
 F_{ab}=-F_{ba} & F^{\alpha}_{\beta}, \quad F^{\alpha}_{\alpha}=0  \\
\hline
 G_{abc}=G_{[abc]}={}^{*}G_{abc}& G_{\alpha\beta}=G_{(\alpha\beta)}  \\
\hline
 G_{abc}=G_{[abc]}=-{}^{*}G_{abc}& G^{\alpha\beta}=G^{(\alpha\beta)}  \\
\hline
  T_{ab}=T_{(ba)},\quad T^{a}_{a}=0 & T_{\alpha\beta,\gamma\delta}=
  T_{\gamma\delta ,\alpha\beta}=T_{[\alpha\beta],[\gamma\delta]},
  \quad T_{[\alpha\beta,\gamma\delta]}=T_{\alpha[\beta,\gamma\delta]}=0 \\
\hline
 \left(S_{a}\right)_{\alpha},\quad \left(\Sigma^{a}S_{a}\right)^{\alpha}=0 &
 S_{\alpha,\beta\gamma}= S_{\alpha,[\beta\gamma]}, \quad S_{[\alpha,\beta\gamma]}=0 \\
\hline
\end{array}$$
In addition we will present some useful formulas and relations
\begin{eqnarray}
  V^{\alpha\beta} &=& \frac{1}{2}\epsilon^{\alpha\beta\gamma\delta}V_{\gamma\delta}
  \quad ,\quad V_{\alpha\beta} = \frac{1}{2}\epsilon_{\alpha\beta\gamma\delta}
  V^{\gamma\delta} \\
   \partial_{[\varepsilon\gamma}V_{\beta\alpha]}&=& \frac{1}{12}
   \epsilon_{\varepsilon\gamma\beta\alpha}\partial^{\mu\nu}V_{\mu\nu}\quad , \quad
   \partial_{[\varepsilon\lambda}\delta^{\delta}_{\mu]}=\frac{1}{3}
   \epsilon_{\mu\varepsilon\lambda\gamma}\partial^{\delta\gamma} \\
\partial_{\mu\nu}\partial^{\lambda\nu} &=&\frac{1}{4}\partial_{\mu\nu}\partial^{\mu\nu}
\delta^{\lambda}_{\mu}\quad , \quad \Box =
-\partial^{a}\partial_{a}=
\frac{1}{4}\partial_{\mu\nu}\partial^{\mu\nu}
\end{eqnarray}
 For the $(1,0)$ case  we used  $SU(2)$ index rules which are presented here
 \begin{eqnarray}
  &&V^{i} =\varepsilon^{ij}V_{j},\qquad  V_{i}
  =V^{j}\varepsilon_{ji}\\
  &&\varepsilon^{ij}\varepsilon_{ik}=\delta^{i}_{k} \quad ,\quad
  \varepsilon^{[ij}V^{k]}=0\\
  &&V^{k(i}\varepsilon^{j)l}-V^{l(i}\varepsilon^{j)k}=
  V^{ij}\varepsilon^{kl},\quad V^{ij}=V^{(ij)}\\
  &&V^{(ij}\varepsilon^{k)l}-V^{(ij}\varepsilon^{l)k}=
  \frac{4}{3}V^{ij}\varepsilon^{kl}
\end{eqnarray}
  Some  formulas for the $(2,0)$ case ($USp(4)$ R-symmetry case)are:
\begin{eqnarray}
  &&V^{i} =\Omega^{ij}V_{j} \qquad ,\qquad  V_{i}
  =V^{j}\Omega_{ji}\\
  &&\Omega_{i[j}\Omega_{kl]}=\frac{1}{3}\epsilon^{ijkl} \quad ,
  \quad \epsilon^{[ijkl}V^{m]}=0 \\
  &&\Omega^{[ij}\Phi^{kl]}=\frac{1}{2}\left(\Omega^{[ij}\Phi^{k]l}
  -\Omega^{l[k}\Phi^{ij]}\right)=0\\
  &&\textnormal{if} \qquad \Phi^{ij}=\Phi^{[ij]} \quad \textnormal{and}
   \quad  \Phi^{ij}\Omega_{ij}=0
\end{eqnarray}
Multiplications of the derivative operators give:
\begin{eqnarray}
   &&D^{i}_{\alpha}D^{j}_{\beta}=\frac{i}{2}\varepsilon^{ij}\partial_{\alpha\beta}+
   \varepsilon^{ij}D_{(\alpha\beta)}+D^{(ij)}_{[\alpha\beta]} \quad \textnormal{for (1,0)}; \\
  &&D^{i}_{\alpha}D^{j}_{\beta}=\frac{i}{2}\Omega^{ij}\partial_{\alpha\beta}+
   \Omega^{ij}D_{(\alpha\beta)}+D^{(ij)}_{[\alpha\beta]} + D^{[ij]}_{(\alpha\beta)}
   ,\\
   &&\Omega_{ij}D^{[ij]}_{(\alpha\beta)}=0 \qquad \quad \textnormal{for
   (2,0)} .\nonumber
\end{eqnarray}

\end{document}